\documentclass[twocolumn,prb,superscriptaddress,showpacs,amsmath,amssymb]{revtex4}
\usepackage{graphicx}% Include figure files
\usepackage{bm}% bold math

\begin{document}
\title{Doping Dependence of Polaron Hopping Energies in La$_{1-x}$Ca$_x$MnO$_3$ ($0\leq x\leq 0.15$)}
\author{K. P. Neupane}
\affiliation{Department of Physics, University of Miami, Coral
Gables, Florida 33124}
\author{J.~L.~Cohn}
\affiliation{Department of Physics, University of Miami, Coral
Gables, Florida 33124}
\author{H. Terashita$^{\ast}$}
\affiliation{Department of Physics, Montana State University,
Bozeman, Montana 59717}
\author{J. J. Neumeier}
\affiliation{Department of Physics, Montana State University,
Bozeman, Montana 59717}

\begin{abstract}
Measurements of the low-frequency ($f\leq 100$~kHz) permittivity at $T\lesssim 160$~K and dc resistivity
($T\lesssim 430$~K) are reported for La$_{1-x}$Ca$_x$MnO$_3$
($0\leq x\leq 0.15$). Static dielectric constants
are determined from the low-$T$ limiting
behavior of the permittivity. The estimated polarizability for
bound holes $\sim 10^{-22}$~cm$^{-3}$ implies a radius comparable
to the interatomic spacing, consistent with the small polaron
picture established from prior transport studies near room
temperature and above on nearby compositions. Relaxation peaks in the dielectric loss
associated with charge-carrier hopping yield activation energies
in good agreement with low-$T$ hopping energies determined from
variable-range hopping fits of the dc resistivity. The doping
dependence of these energies suggests that the orthorhombic,
canted antiferromagnetic ground state tends toward
an insulator-metal transition that
is not realized due to the formation of the ferromagnetic insulating
state near Mn$^{4+}$ concentration $\approx 0.13$.

\end{abstract}
 \maketitle
\section{\label{sec:Intro} Introduction}

The Mn$^{3+}$-rich insulating compounds, La$_{1-x}$A$_x$MnO$_3$
(A=Sr, Ca, $x\leq 0.15$) have attracted considerable interest in
recent years for their inhomogeneous magnetic ground
states,\cite{BiotteauHennion} and complex structural phase
diagrams,\cite{LSMOStructure,Pissas} believed to be consequences
of competing magnetic, lattice, and Coulomb
interactions.\cite{Dagotto} With regard to transport properties,
there is general consensus that the charge-carriers in the
high-$T$ paramagnetic phases of these compounds are small
polarons.\cite{Jaime,DeTeresa,CohnReview}  However, few systematic
studies of the doping dependence of polaron transport in the
insulating regime have been reported.\cite{LSMOPolarons}  The experimental situation is
complicated by the sensitivity of the structural and magnetic
properties to deviations in stoichiometry\cite{LSMOStructure,Pissas} and the
uncertain stoichiometry of specimens employed in many early transport studies.

Here we present dielectric and dc resistivity measurements on a series of La$_{1-x}$Ca$_x$MnO$_3$
polycrystalline specimens with $x\leq 0.15$ that offer new insight into the role of electron correlations
on the changes in ground state magnetic and crystal structure that occur\cite{BiotteauHennion,LSMOStructure,Pissas}
near Mn$^{4+}$ content, $p=p_c\simeq 0.13$.  Consistent with the small-polaron picture, the doping dependence of the
dielectric constant yields a bound-polaron polarizability that implies a localization length comparable to a
lattice spacing.  In agreement with prior work\cite{LSMOPolarons} on La$_{1-x}$Sr$_x$MnO$_3$
we find that the high-temperature polaron hopping energy is independent of $p$ up to
$p_c$, above which it begins to decrease.  Particularly interesting are low-temperature measurements
of dielectric relaxation and dc resistivity which indicate a polaron hopping energy that tends toward zero
near $p_c$.  These results suggest that charge-carriers within the orthorhombic, canted
antiferromagnetic (CAF) phase are approaching an insulator-metal transition that is not achieved due to the
emergence of the monoclinic and ferromagnetic insulating (FMI) phase.

\section{\label{sec:Expt}Experiment}

Polycrystalline specimens of  La$_{1-x}$Ca$_x$MnO$_{3+\delta}$ were prepared by standard solid-state
reaction under Ar atmosphere; the preparation methods, structural characterization and magnetic properties
are reported elsewhere.\cite{TerashitaNeumeier}
Powder x-ray diffraction revealed no secondary phases; the average
Mn$^{4+}$ (hole) concentration ($p=x+2\delta$) was determined by iodometric titration (see Table I).
A small and nearly $x$-independent oxygen excess ($\delta_{av}\approx 0.017$) in the present specimens
is attributed to cation vacancies\cite{VanRoosmalen,Dabrowski} which contribute additional holes
uniformly,\cite{Dabrowski} similar to Ca substitution.  The structural and magnetic properties of
the present specimens are consistent with those of stoichiometric compounds.\cite{Dabrowski,Pissas}

AC impedance measurements were conducted with an HP4263B LCR meter
at frequencies $f=$~100~Hz, 120~Hz, 10~kHz, 20~kHz, 100~kHz using
a 4-terminal pair arrangement. Reliable measurements of
$\varepsilon$ were restricted to $T\lesssim 160$ K where the
capacitive reactance was sufficiently large ($\gtrsim 0.1\Omega$).
Typical specimen dimensions were $3\times 1.0 \times 0.5$ mm$^3$.
Silver paint electrodes were applied on the largest, polished
faces of the specimens.  Contact capacitance can lead to
apparently large values\cite{Lunkenheimer} of $\varepsilon$ and
thus some care is required to distinguish the true response of the
sample. To rule out the influence of contacts, the impedances of
several specimens were remeasured after further polishing to
reduce the electrode spacing by at least a factor of two; in all
cases the low-temperature data used to determine $\varepsilon_0$
agreed within geometric uncertainties of $\pm 10$\%. The results were also independent of
applied ac voltage in the range 50mV-1V.  At higher temperatures
where dielectric relaxation was observed, some dependence on the electrode
spacing was evident, consistent with a contribution from contact capacitance.
This issue is addressed further below.  Four-probe dc resistivity measurements
were performed in separate experiments in air (high-$T$) and vacuum (low-$T$)
with current applied along the long axis of the same specimens used for ac impedance.

\section{\label{sec:ResultsDisc} Results and Analysis}

Quite generally, the complex dielectric permittivity of a solid,
$\varepsilon=\varepsilon^{\prime}-i\varepsilon^{\prime\prime}$,
can be expressed as,
$\varepsilon=\varepsilon_{\infty}+\varepsilon_l+\varepsilon_d$.
$\varepsilon_{\infty}$ is the high-frequency dielectric constant
associated with displacements of ionic charge distributions
relative to their nuclei.  The lattice contribution,
$\varepsilon_l$, arises from displacements of ions and their
charge distributions. $\varepsilon_d$ represents a dipolar
contribution, associated in the present materials with
charge-carrier hopping as discussed below.  $\varepsilon_{\infty}$ and
$\varepsilon_l$ are generally frequency- and temperature-independent at low $T$.

\begin{figure}
\includegraphics[width = 3.8in,clip]{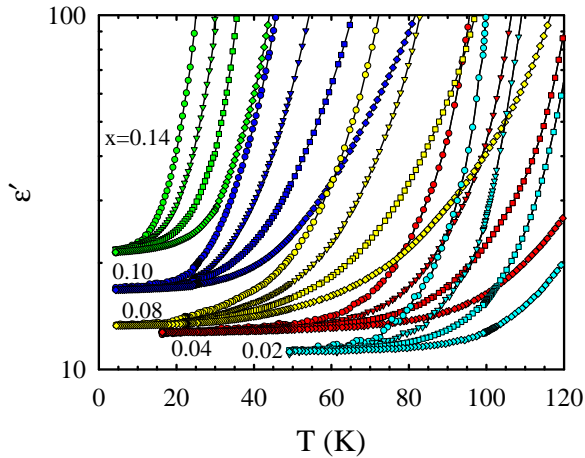}%
\vglue -.3in
\caption{(Color online) $\varepsilon^{\prime} ${\it vs} $T$ for La$_{1-x}$Ca$_x$MnO$_3$ specimens at frequencies
100 Hz, 1 kHz, 10 kHz, 100 kHz (increasing from left to right for each composition).}
\label{LCMOepsvsT}
\end{figure}

\begin{figure}
\includegraphics[width =3.2in,clip]{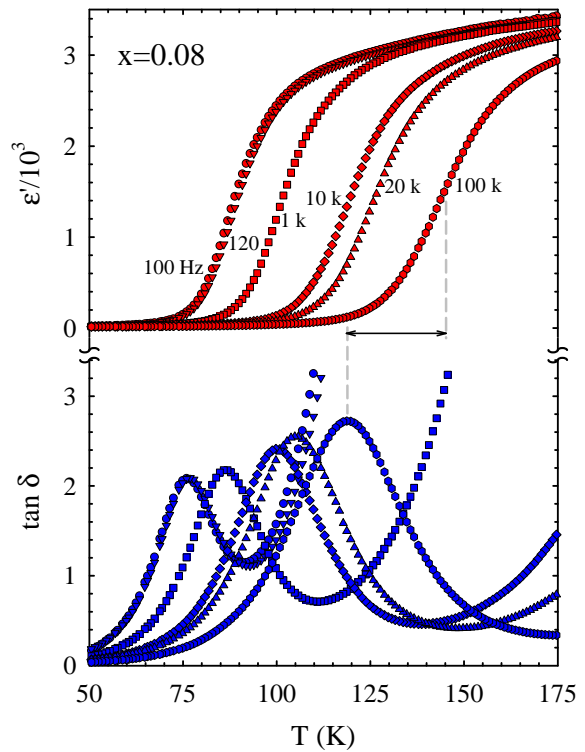}%
\vglue -.2in
\caption{(Color online) $\varepsilon^{\prime}$ and loss tangent {\it vs} $T$ at different
frequencies for x=0.08.}
\vglue -.2in
\label{LCMORelax}
\end{figure}

Figure~\ref{LCMOepsvsT} shows $\varepsilon^{\prime}(T)$ at four
frequencies for five Ca concentrations in the low-$T$ regime where
$\varepsilon^{\prime}$ becomes frequency and temperature
independent.  The intrinsic static dielectric constant
$\varepsilon_0\equiv\varepsilon^{\prime}(T\to 0)$ can be
determined directly from the data. The sharp increase of
$\varepsilon^{\prime}(\omega)$ and onset of frequency dispersion
with increasing temperature, signatures of dielectric relaxation,
are more clearly manifested in the broader temperature range
of Fig.~\ref{LCMORelax} (for $x=0.08$) as steps in $\varepsilon^{\prime}$ and
maxima in the loss factor, $\tan\delta\equiv \varepsilon^{\prime\prime}/\varepsilon^{\prime}$.

A significant feature of the relaxation behavior for all specimens is that the temperatures of the
maxima in $\tan\delta$ (denoted $T_{max}$) and inflections in $\varepsilon^{\prime}(T)$
differ substantially (dashed lines in Fig.~\ref{LCMORelax}).  This rules out an intrinsic dipole response,
for which these features occur at the same temperature.\cite{Volger,Grubbs}  Schottky barriers are to be expected at
metal-semiconductor contacts with manganites.\cite{Lunkenheimer,Freitas}  The associated depletion layers behave as a
large capacitance in parallel with a resistor, in series with the bulk specimen. Assuming that the
contact capacitance is nearly temperature independent, the time constant of the system ($\tau$) is controlled by the
temperature-dependent bulk resistance of the specimen,\cite{Lunkenheimer} with maxima in $\tan\delta$ occurring for $\omega\tau\simeq 1$.
Consistent with this picture, Freitas {\it et al.}\cite{Freitas} have demonstrated for a related manganite compound
that large values of $\varepsilon^{\prime}$ (at $T\geq 150$~K in Fig.~\ref{LCMORelax}) are sensitive to the type of
contacts employed (e.g. silver paint, vapor-deposited metals), but that the charge-carrier activation energies
derived from the $\tan\delta$ maxima are contact independent.
\vglue -.15in

\begin{figure}
\includegraphics[width =4.in,clip]{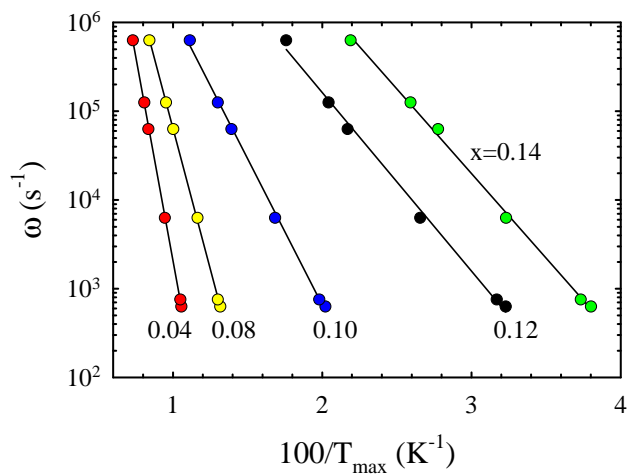}%
\vglue -.2in
\caption{(Color online) $\omega$ vs 100/$T_{max}$ for several Ca concentrations.
Solid lines are linear least-squares fits.}
\label{TmaxvsOmega}
\vglue -.2in
\end{figure}

Figure~\ref{TmaxvsOmega} shows a semilog plot of $\omega$ {\it vs} $100/T_{max}$ for several specimens, demonstrating
Arrhenius behavior for the relaxation rates, $\tau^{-1}(T)=\omega_0\exp(-E_{\delta}/k_BT)$.  Values of $E_{\delta}$
and $\omega_0$ were determined from the slopes and intercepts, respectively, of
linear-least-squares fits (solid lines, Fig.~\ref{TmaxvsOmega}) and are listed in Table I.
% Table
\begin{table*}
\caption{\label{tab:Table1} Nominal Ca and measured Mn$^{4+}$
concentrations, dielectric constant, activation energies ($E_{\delta}$) and prefactors ($\omega_0$) of
Arrhenius relaxation rates determined from tan$\delta$, and $T_0^{1/4}$ and $E_{\rho/T}$
determined from the dc resistivities for La$_{1-x}$Ca$_x$MnO$_3$ specimens.}
\begin{ruledtabular}
\begin{tabular}{ccccccc}
 $x$& $p\equiv$~Mn$^{+4}$~(\%)&$\varepsilon_0$&$E_{\delta}$~(meV)&$\omega_0$~(s$^{-1}$)&$T_0^{1/4}~({\rm K}^{1/4})$&$E_{\rho/T}$~(meV)\\
\hline
0.00&4.0&17.6&125&1.49$\times 10^{11}$&{--}&{--}\\
0.02&5.7&11.3&202&2.03$\times 10^{13}$&234& 231  \\
0.04&7.8&12.7&182&3.26$\times 10^{12}$&228& 232  \\
0.06&9.9&11.7&155&1.09$\times 10^{12}$&208& 233  \\
0.08&11.3&13.8&126&1.42$\times 10^{11}$&185&233  \\
0.10&12.6&16.3&65.5&2.58$\times 10^{9}$&155&233  \\
0.12&14.1&17.8&43.9&4.42$\times 10^{9}$&148&220  \\
0.14&17.4&21.4&37.9&1.05$\times 10^{10}$&134&209 \\
0.15&{--}&20.9&26.8&5.21$\times 10^{8}$& {--}&{--} \\
\end{tabular}
\end{ruledtabular}
\end{table*}

\begin{figure}
\includegraphics[width =3.3in,clip]{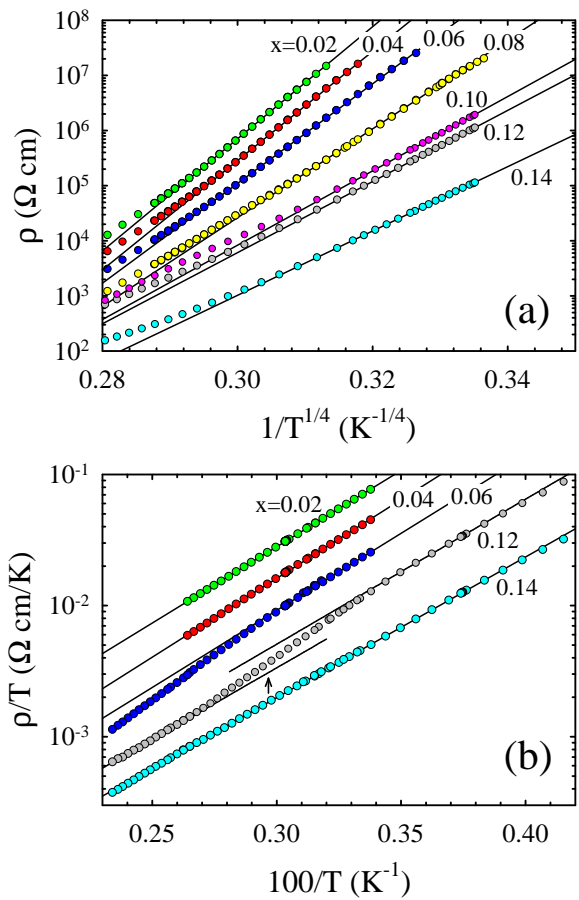}%
\caption{(Color online) (a) $\rho$ vs 1/$T^{1/4}$ at low temperatures for La$_{1-x}$Ca$_x$MnO$_3$,
(b) high-temperature data plotted in the adiabatic small-polaron form. Data for $x=0.08$ and 0.10 are omitted for
clarity in (b).  Solid lines in both figures are linear least-squares fits.}
\label{RofT}
\vglue -.2in
\end{figure}

Figure~\ref{RofT} shows dc resistivity measurements for the same specimens. The low-$T$ data ($T\leq 160$~K) in Fig.~\ref{RofT}~(a)
are plotted against $T^{-1/4}$, suggesting the variable-range hopping form,\cite{MottDavis}
$\rho=\rho_0\exp[-(T_0/T)^{1/4}]$. The $T^{-1/2}$ Efros-Shklovskii\cite{EfrosShklovskii} hopping form is
equally satisfactory over the limited temperature range of the linear least-squares fits (solid lines).
High-$T$ resistivity data ($240~{\rm K}\leq T\leq 440$~K) in Fig.~\ref{RofT}~(b) are shown plotted to emphasize the adiabatic small-polaron form
widely established for the paramagnetic phase of these materials,\cite{Jaime,DeTeresa,CohnReview,LSMOPolarons}
$\rho/T=A\exp(E_{\rho/T}/k_BT)$.  Solid lines are linear least-squares fits used to determine $E_{\rho/T}$; these
values and those for $T_0^{1/4}$ are listed in Table I.
The Jahn-Teller (JT) structural transition is clearly evidenced by a vertical displacement of the data for
$x=0.12$ [arrow in Fig.~\ref{RofT}~(b)] over a broad temperature interval over which the lattice constants are
observed to change.\cite{Pissas}  The slopes above and below
the transition are the same, in agreement with similar studies of La$_{1-x}$Sr$_x$MnO$_3$.\cite{LSMOPolarons}
A transition for the $x=0.06$ specimen is also evident, but is not completed at the maximum temperature investigated.
The JT transition for the $x=0.14$ specimen is expected somewhat below room temperature\cite{Pissas}; the absence of
a noticeable feature in $\rho(T)$ is consistent with the systematic broadening and eventual disappearance of features
in structural and magnetic measurements with increased doping near this hole concentration.\cite{Pissas}

\section{\label{sec:Discussion} Discussion}

\begin{figure}
\includegraphics[width =3.6in,clip=]{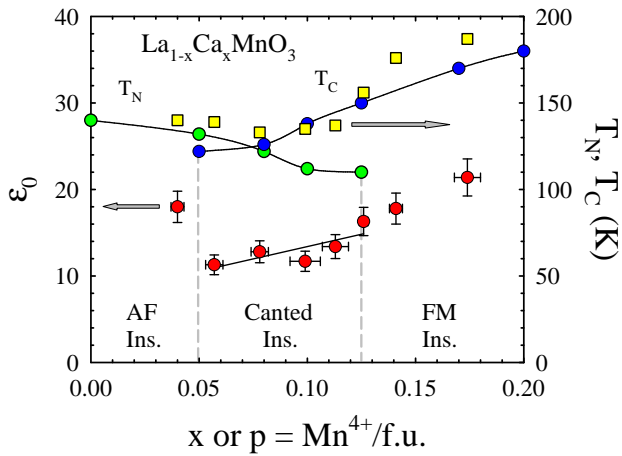}%
\caption{(Color online) $\varepsilon_0$ {\it vs} Mn$^{4+}$ concentration from the present study (left ordinate).
The solid line is a linear least-squares fit to the data for $0.057\leq p\leq 0.126$.
Also shown are data for $T_N$ and $T_C$ (right ordinate) determined from magnetization measurements on
polycrystalline specimens from the same batch as those used here for dielectric measurements
(squares, Ref.~\protect\onlinecite{TerashitaNeumeier}), and from neutron diffraction on single crystals
(circles, Ref.~\protect\onlinecite{BiotteauHennion}).  The vertical dashed lines delineate approximate
low-$T$ magnetic phase boundaries from Ref.~\protect\onlinecite{BiotteauHennion}.}
\label{LCMOepsvsx}
\end{figure}

The dependence of $\varepsilon_0$ on the hole concentration, $p$,
is shown in Fig~\ref{LCMOepsvsx}.  For the nominally undoped
compound\cite{CohnPeterca} LaMnO$_3$, $\varepsilon_0=18$.  A
smaller value, $\varepsilon_0\approx 12$, is observed for $p\sim 0.05-0.10$, and
an approximately linear increase for $p\geq 0.10$. Also
shown are the magnetic phase boundaries defined by $T_N$ or $T_C$
from magnetization data on polycrystals from the same batches as
those reported here\cite{TerashitaNeumeier} (squares and right
ordinate, Fig.~\ref{LCMOepsvsx}) and neutron diffraction studies
of single crystals\cite{BiotteauHennion} (circles and right
ordinate, Fig.~\ref{LCMOepsvsx}).  The recently clarified
structural phase diagram\cite{Pissas} indicates orthorhombic and
monoclinic structures for the CAF and FMI regimes, respectively,
with a transition occurring in the range $0.12\leq p\leq 0.14$.

Let us first consider the decrease in $\varepsilon_0$ at the lowest hole concentrations.  This behavior is
contrary to expectations based on the behavior of conventional semiconductors\cite{SiEps} where
the dielectric constant of the (undoped) host lattice
($\varepsilon_{0,h}$) is enhanced at low temperatures by the presence of electrons or holes bound to
donor or acceptor impurities, respectively.  Local fields produced by charged defects, La and Mn vacancies,
and excess oxygen,\cite{VanRoosmalen,Dabrowski} can influence the polarizability associated with bound
holes in the ground state.  The observed decrease in
$\varepsilon_0$ by $\sim 30$~\% would require a corresponding decrease in the acceptor polarizability
due to differences in the long-range magnetic and/or orbital structure of the A-type AF and CAF states.

The increase of $\varepsilon_0$ for $p\geq 0.05$ is consistent with a
growing contribution from holes bound to acceptors in the present materials (Ca or excess oxygen).
An estimate of the polarizability can be found from
the solid line in Fig.~\ref{LCMOepsvsx} using
the dilute limit of the Castellan-Seitz\cite{CastellanSeitz} expression,
$\varepsilon_0-\varepsilon_{0,h}=4\pi N\alpha_0$, where $N\equiv p/V_{fu}$ is the acceptor density
and $V_{fu}$ is the volume per formula unit.\cite{TerashitaNeumeier}
This yields $\alpha_0\simeq 2.8\times 10^{-22}\ {\rm cm}^3$ which implies (for hydrogenic impurities)
a Bohr radius $a_H=(2\alpha_0/9\varepsilon_{0,h})^{1/3}\simeq 1.7{\rm\AA}$ (we take $\varepsilon_{0,h}=12$ from
Fig.~\ref{LCMOepsvsx} as representative of the CAF phase).  Thus the $\varepsilon_0$ data imply
electronic states for the bound holes that are localized at individual atomic sites, consistent with general consensus from
transport studies near room temperature and above, that doped holes move as small polarons.\cite{Jaime,DeTeresa,CohnReview}

\begin{figure}
\vglue -.12in
\includegraphics[width =3.4in,clip=]{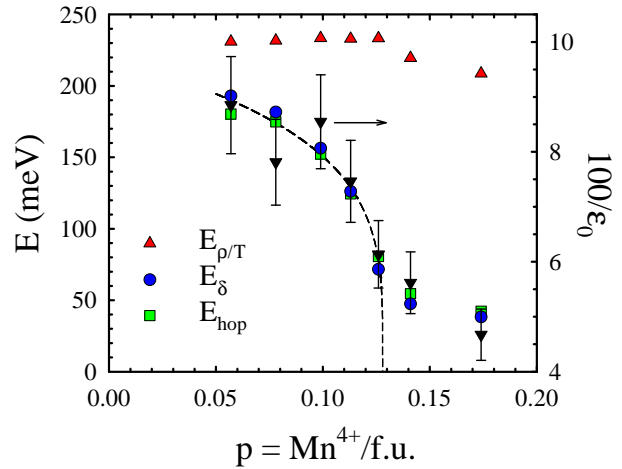}%
\caption{(Color online) Activation energies (left ordinate) determined from dielectric
relaxation ($E_{\delta}$) and dc resistivities at high-$T$ ($E_{\rho/T}$) and low-$T$ ($E_{hop}$),
and inverse dielectric constant (right ordinate) {\it vs} Mn$^{4+}$ concentration.}
\label{PolaronEnergies}
\end{figure}

It is instructive to examine and compare the various polaron energies determined from our measurements.
As noted above, we expect $E_{\delta}$ to represent the low-$T$ charge-carrier hopping energy.  Supporting
this conclusion we find excellent agreement between $E_{\delta}$ and the hopping energy determined from the
dc resistivity variable-range hopping fits,\cite{MottDavis} $E_{hop}=0.25k_BT_0^{1/4}T^{3/4}$, where $T$ in this
expression was taken as the average $T_{max}$ for each composition.  Figure~\ref{PolaronEnergies} shows these energies
as a function of doping.  Also plotted is the high-$T$ polaron activation energy,
$E_{\rho/T}$, which remains approximately constant at $\sim 230$~meV up to $p_c\approx 0.13$.  At higher doping
it decreases, a trend quite similar to that reported for La$_{1-x}$Sr$_x$MnO$_3$.\cite{LSMOPolarons}
Most striking is the behavior of $E_{hop}$ which appears to be dropping sharply toward zero at $p_c$, and exhibits a
clear change in doping dependence for $p>p_c$.  The dashed curve in
Fig.~\ref{PolaronEnergies} is an empirical power-law fit, $E_{hop}=220(1-p/0.128)^{1/4}$.
The inverse dielectric constant, also shown (right ordinate, Fig.~\ref{PolaronEnergies}),
appears to follow a very similar behavior.

$E_{hop}(p)$ is reminiscent of the behavior of conventional doped semiconductors\cite{Shafarman} near the insulator-metal transition (IMT),
with $p_c$ the critical density.  This point of view is encouraged by the observation that $p_c$ and $a_H$ are close to satisfying the
Mott criterion\cite{MottCriterion} for an IMT, $N_c^{1/3}a_H\simeq 0.26\pm 0.05$.  Experimentally we find $\approx 0.22$.
Thus the data suggest that the ground-state structural and magnetic transitions\cite{BiotteauHennion,Pissas}
occurring near $p_c$ are a consequence of the rapidly growing charge-carrier kinetic energy as $p\to p_c$ within
the orthorhombic, CAF phase.  The change in orbital-ordering scheme in the monoclinic, FMI phase\cite{Pissas} might be
viewed as a compromise among competing magnetic, lattice and Coulomb energies that suppresses the rate of growth in kinetic energy with
doping, and ``delays'' the IMT to higher Mn$^{4+}$ concentration.  Given this low-$T$ picture of proximity to an IMT,
it is natural to hypothesize that the decrease in the high-$T$ small-polaron energy, $E_{\rho/T}$, at $p>p_c$
reflects the cooperative effects of overlapping polarons.

Scaling behavior is to be expected near the IMT, with physical properties controlled by a localization length
diverging as $\xi=\xi_0(1-p/p_c)^{-\nu}$, with $\nu$ a critical exponent.  With\cite{MottDavis} $T_0\propto 1/\xi^3$,
this yields $T_0^{1/4}\propto (1-n/n_c)^{3\nu/4}$, and our empirical fit (dashed curve, Fig.~\ref{PolaronEnergies})
implies $\nu\simeq 1/3$.  Though smaller than values in the range $1/2\leq\nu\leq 1$ that have been reported for
doped Si and Ge,\cite{Shafarman,CritExpsSiGe} this estimate of $\nu$ should be viewed with caution as a lower bound
since the scaling theory applies only very close to $p_c$ where the present data are not sufficiently dense.
Furthermore, the inhomogeneous ground state of these materials may alter scaling behavior.  This is especially
likely near $p_c$ where both orthorhombic and monoclinic domains may coexist\cite{Pissas} and
where the coalescence of ferromagnetic clusters of nanoscopic size embedded in the CAF matrix
have been deduced from neutron scattering studies.\cite{BiotteauHennion}

In summary, combined dielectric and resistivity measurements reveal new systematics of electronic degrees
of freedom in the complex magnetic and structural phase behavior of lightly doped La$_{1-x}$Ca$_x$MnO$_3$.
The doping dependence of the dielectric constant and low-$T$ hopping energies suggest that adding holes within the
CAF ground state drives the electronic system toward an insulator-metal transition
near Mn$^{4+}$ concentration $\approx 0.13$.  That this transition isn't realized is evidently a manifestation of
competition from magnetic and lattice energies reflected in the emergence of the FMI phase.

\section{\label{sec:Ack} Acknowledgments}

The authors acknowledge experimental assistance from Mr. C. Chiorescu.
This material is based upon work supported by the National Science Foundation under grants
DMR-0072276 (Univ. Miami) and DMR-0504769 (Montana State Univ.), and by an award from
the Research Corporation (Univ. Miami).

\noindent
$^{\ast}$ Present address: Research Center J\"ulich, D-52425 J\"ulich

\newpage
\vfill
\newpage
%-------------------------Figures
\end{document}